===== MOLECULAR SPECTROSCOPY =====

# Modeling of the Time-Resolved Vibronic Spectra of Polyatomic Molecules: The Formulation of the Problem and Analysis of Kinetic Equations

**S. A. Astakhov and V. I. Baranov**

*Institute of Geochemistry and Analytical Chemistry, Russian Academy of Sciences, Moscow, 117975 Russia*
*e-mail: spec.lab@g23.relcom.ru*
Received March 28, 2000

**Abstract**—A semiempirical parametric method is proposed for modeling three-dimensional (time-resolved) vibronic spectra of polyatomic molecules. The method is based on the use of the fragment approach in the formation of molecular models for excited electronic states and parametrization of these molecular fragments by modeling conventional (one-dimensional) absorption and fluorescence spectra of polyatomic molecules. All matrix elements that are required for calculations of the spectra can be found by the methods developed. The time dependences of the populations of a great number ($>10^3$) of vibronic levels can be most conveniently found by using the iterative numerical method of integration of kinetic equations. Convenient numerical algorithms and specialized software for PC are developed. Computer experiments showed the possibility of the real-time modeling three-dimensional spectra of polyatomic molecules containing several tens of atoms. © *2001 MAIK "Nauka/Interperiodica".*

## INTRODUCTION

Recently, considerable advances have been made in the experimental methods of molecular vibronic spectroscopy. On the one hand, a very high resolution ($\Delta\nu \approx 1$ cm$^{-1}$) of the vibrational structure in the spectra was achieved in supersonic jets, and, on the other, modern laser methods provide the observation of dispersive spectra (fluorescence from single excited vibronic levels) and time-resolved dynamic spectra of molecules excited by short (pico- and femtosecond) light pulses (see, for example, [1–11]).

The modern vibronic spectroscopy is developing in the direction of studies of the time-resolved line spectra. This necessitates the development of the corresponding theory, computational methods, and the construction of molecular models for the calculation, interpretation, and prediction of such spectra and obtaining from them information on the properties of molecules.

The conventional absorption and fluorescence spectra of polyatomic molecules have long been analyzed theoretically. The theory was developed from the simplest methods, which used various generalized spectral parameters (see, for example, [12]), to the methods based on the fundamental molecular model and its parameters, which were not directly related to the spectral experiment (potential surfaces of the ground and excited states in natural molecular coordinates) [13]. The advantage of this model is obvious, because it allows one to describe not only spectral but also other physicochemical properties of polyatomic molecules (even, for example, the development of chemical reactions [14, 15]). Within the framework of the parametric theory of vibronic spectra, calculation methods were developed which allow one to construct the spectral representation of the molecular model specified by a set of parameters [16–19]. The second, no less important problem, has also been solved, namely, the determination of parameters of the potential surfaces of molecules in excited states [20].

This provided the basis for the formulation and solution of the problems of modeling and prediction of one-quantum, absorption, and fluorescence vibronic spectra. It seems reasonable to develop this parametric approach for modeling the time-resolved spectra, which obviously can give new information compared to conventional vibronic spectra. However, at first glance, to calculate the time-resolved spectrum of a complex molecule, it is sufficient to know all the probabilities $w_{ij}$ of vibronic transitions, which can be easily calculated by the parametric method; nevertheless, the modeling of such spectra involves a number of problems in practice.

First, it is not clear to what degree the parameters of a molecular model that were obtained (calibrated) for the spectra of a certain type will be appropriate for a quantitative modeling of the spectra of another type (time-resolved spectra).

Second, the influence of a medium (intermolecular interactions) can substantially change the emission of excited molecules. The problem arises as to how to separate the contributions from nonradiative and radiative transitions to the time-resolved spectra.

Third, processes of absorption and radiative and nonradiative relaxation in excited molecules can occur not only inside the same molecular isomer but can result in the formation of other isomers (which is, in particular, of great interest for photochemistry). Therefore, it is necessary to determine the probabilities of optical and nonradiative transitions between molecular isomers.

Fourth, a purely computational problem arises. In addition to the calculation of many probabilities $w_{ij}$ of vibrational and vibronic transitions in the determination of time dependences of level populations $n_i(t)$, it is necessary to solve a system of linear differential equations of a very high dimensionality ($>10^3$) even for molecules of a moderate size containing $\approx$20–30 atoms. Although the general methods are well known, the applicability of the corresponding computational procedures for modeling spectra is not obvious and requires a special analysis (because of a high dimensionality of the problem and the fast computations required).

At present, all these questions have no clear answer, although studies are being performed in some fields (for example, nonradiative transitions [21], probabilities of optical transitions between isomers [13, 14], and generalized inverse vibronic problems [22]). This paper is devoted to the first part of the general problem, namely, the solution of the direct problem of calculation of the time-resolved vibronic spectrum for real large molecular systems. We restrict ourselves to the case of an isolated molecule and transitions inside only one molecular isomer.

## PARAMETRIC APPROACH

An adequate choice of the parameters of a molecular model is very important in the solution of a direct spectral problem.

Note that the excited-state lifetimes, decay rates for individual lines in the spectrum, and the quantum yield of fluorescence are the key parameters (along with frequencies and intensities of vibronic lines) in the analysis and modeling of time-resolved spectra [23, 24]. These quantities, which determine the time dependence of the fluorescence intensity and can be obtained directly from experiments, are in fact the parameters of a specific experiment rather than the parameters of the molecular model. Knowing these parameters, one can reproduce the experiment numerically, but it is impossible to predict the behavior of another or the same molecule under different conditions. Such an approach is incapable of predicting physicochemical properties of molecules and is not constructive in this sense for the development of methods of modeling time-resolved vibronic spectra of polyatomic molecules.

We suggest using a parametric method, which was developed earlier for calculations of the conventional (not time-resolved) vibronic spectra of polyatomic molecules and is based on the fragmentary method of formation of molecular models and a special system of parameters possessing all the required properties [13, 20, 25]. The essence of the fragmentary approach is that the calibration of the values of parameters of the molecular fragments can be performed for the simplest molecules and then these fragments can be used for constructing models of more complex compounds and predicting their spectra. It is important that both the prediction of spectral properties and the development of a system of parameters of model fragments are preformed in the same experiments (absorption or fluorescence).

The main idea of this approach is to calibrate the values of parameters of the molecular model (or fragment) in the simplest experiment (one-dimensional absorption or fluorescence spectrum) and then to predict the results of another, more complex, time-resolved experiment, which describes in detail the dynamics of excited states and relaxation processes in polyatomic molecules. This will allow one to use data bases for molecular fragments [20, 25] in computer experiments for direct calculations of multidimensional spectra, which can be directly compared with the corresponding real experiments, and to refine the parameters of the molecular model. The computer experiment devoted to the study of the effect of various factors on the shape of the time-resolved spectrum can be also used instead of real experiments, which often require more complex equipment and are time-consuming.

It is clear that such an approach for modeling multidimensional vibronic spectra can be realized in principle. The main problem, which determines the predictive ability of this method, is the degree of transferability of the parameters of molecular fragments used. As for any parametric method, the answer to this question can be obtained from model calculations of sufficiently representative series of molecules and comparison of the results with experiments (this question will be discussed elsewhere). Nevertheless, one can already expect the high efficiency of these models, taking into account their good predictive ability in modeling conventional IR and UV spectra of polyatomic molecules [20, 25–27].

Along with a choice of the molecular model, it is necessary to solve the computational problems involved. The time-resolved vibronic fluorescence spectrum can be represented as a three-dimensional surface—the dependence of the intensity on the emission frequency and the observation time. To construct such a surface, one should calculate the time dependence of the intensity of each spectral line, which is proportional to the probability $w_{ij}$ of the corresponding vibronic transition from the state $i$ to the state $j$ and the population $n_i$ of the level from which this transition occurs:

$$I_{ij}(\nu_{ij}, t) = h\nu_{ij}w_{ij}n_i(t). \tag{1}$$

Therefore, to calculate a three-dimensional spectrum for a specified molecular model, it is necessary, first, to determine the probabilities of vibrational and vibronic transitions between all energy levels of the molecule and, second, by integrating kinetic equations, to obtain the time dependences of populations of all levels and intensities of spectral lines. Note that the number of vibronic states, which affect the intensity $I_{ij}(\nu_{ij}, t)$ of spectral lines, can be very large for polyatomic molecules ($>10^3$).

The methods for calculating the probabilities $w_{ij}$ of vibrational and vibronic transitions are developed in detail, including the corresponding software [13, 16–19, 26, 28]. The second problem requires the development of the efficient methods for solving kinetic equations, because preliminary estimates showed that the simplest methods could not be applied to real time computer experiments with polyatomic molecules due to a great dimensionality of such systems. Let us discuss this question in more detail.

## ANALYSIS OF KINETIC EQUATIONS

Let all the excited states of a molecule be numbered from 1 to $N$ in accordance with their increasing energy. Then, the kinetic equations for populations $n_i(t)$ of the energy levels, which represent linear first-order differential equations with constant coefficients $w_{ij}$ equal to the probabilities of transitions from the state $j$ to the state $i$, will have the form

$$\frac{dn_i}{dt} = -w_i n_i + \sum_{j=i+1}^{N} w_{ji} n_j, \quad i = 1, 2, \ldots N, \quad (2)$$

where $w_i = \sum_{j=1}^{i-1} w_{ij} + w_{i0}$ are the total probabilities of transitions from the state $i$ to all low-lying excited states and the ground state ($w_{i0}$). The initial conditions $\{n_i(0), i = 1, 2, \ldots, N\}$ can be different depending on the type of excitation of a molecule (resonance, broadband excitation, etc.).

Consider the possible ways of solving this problem, taking into account that the system of kinetic equations (2) is written for all vibronic states of the molecule, whose number is large ($N > 10^3$), and in modeling multidimensional spectra, especially in solving inverse problems, the calculation should be performed repeatedly for different parameters of the molecular model, probabilities $w_{ji}$, initial conditions, time intervals, etc. Therefore, the method of solving kinetic equations should be capable of performing fast calculations of quite complicated systems.

The system (2) can be written in the matrix form for the population vector $\|n(t)\|$ with components $n_1(t), \ldots, n_N(t)$ as

$$\frac{d}{dt}\|n(t)\| = P\|n(t)\|, \quad (3)$$

where $P$ is the upper triangle probability matrix with elements

$$p_{ij} = \begin{cases} -w_i, & i = j \\ w_{ji}, & i < j \\ 0, & i > j \end{cases}$$

(its eigenvalues coincide with diagonal elements $w_i$).

In the most general case, the solution of the system of equations (3) is expressed in terms of the exponential of the matrix $P$ and initial conditions $\|n(0)\|$ [29, 30]:

$$\|n(t)\| = e^{Pt}\|n(0)\|. \quad (4)$$

To determine $\|n(t)\|$ with the required accuracy by calculating the exponential in the form of a series $e^{Pt} = \sum_{m=0}^{\infty} \frac{(Pt)^m}{m!}$, one should perform a great number of operations of matrix multiplication for each value of $t$, which represents a cumbersome computational problem and cannot be realized in modeling of the multidimensional spectra of polyatomic molecules with $N > 10^3$.

However, one can solve kinetic equations analytically, which substantially simplifies calculations. First we consider the case when the matrix $P$ does not have multiple eigenvalues.

Let us successively exclude unknowns and represent the solutions of equation (2) in the form

$$n_i(t) = \left(n_i(0) + \sum_{k=i+1}^{N} b_{ki}\right) e^{-w_i t} - \sum_{k=i+1}^{N} b_{ki} e^{-w_k t}. \quad (5)$$

Expressions for coefficients $b_{ki}$ can be obtained by substituting in the equation for an arbitrary $i$th level in system (2) the solutions (5) for higher-lying levels ($n_j(t)$, $j > i$):

$$\frac{dn_i}{dt} = -w_i n_i + w_{i+1\,i} A_{i+1} e^{-w_{i+1} t}$$
$$+ \sum_{k=i+2}^{N} \left(w_{ki} A_k - \sum_{j=i+1}^{k-1} w_{ji} b_{kj}\right) e^{-w_k t},$$

where $A_m = n_m(0) + \sum_{l=m+1}^{N} b_{lm}$. This yields the recurrence relations for $b_{ki}$

$$b_{i+1\,i} = \frac{w_{i+1\,i}A_{i+1}}{w_{i+1} - w_i}$$

$$b_{ki} = \frac{w_{ki}A_k - \sum_{j=i+1}^{k-1} w_{ji}b_{kj}}{w_k - w_i} \quad (6)$$

$$k = i+2, i+3, \ldots, N,$$

where calculations begin from $b_{N\,N-1}$, for which $A_N = n_N(0)$.

Expressions (5) and (6) are convenient for the analysis and calculation of $n_i(t)$ and time dependences of the spectral line intensities; however, they are substantially restricted by the condition $w_k \neq w_i$, which is unacceptable for the methods of modeling the spectra of polyatomic molecules (see below). The solution in the presence of multiple eigenvalues of the matrix $P$, i.e., for the same total probabilities of transitions from some levels ($w_k = w_i$), can be similarly represented as a sum of the products of exponentials and polynomials in $t$ with the corresponding recurrent relations for the coefficients; however, it is cumbersome. This solution can be more easily obtained directly in the matrix form by the transformation of the probability matrix $P$ to the canonical Jordan form [31].

Let the eigenvalues $\lambda_1, \ldots, \lambda_q$ be different among $N$ eigenvalues of the matrix $P$ and the rest $s$ eigenvalues $\lambda_{q+i}$ ($i = 1, 2, \ldots, s$) be multiple with the multiplicity $r_i$. Then, the similarity transformation $M^{-1}PM = J$ reduces the matrix $P$ to the canonical Jordan form $J = \text{diag}\{J_0, J_1, \ldots, J_s\}$, where $J_0$ is the diagonal matrix with elements $\lambda_1, \ldots, \lambda_q$ and $J_i$ are Jordan cells of the order $r_i$

$$J_i = \begin{Vmatrix} \lambda_{q+i} & 1 & 0 & \ldots & 0 & 0 \\ 0 & \lambda_{q+i} & 1 & \ldots & 0 & 0 \\ . & . & . & \ldots & . & . \\ 0 & 0 & 0 & \ldots & \lambda_{q+i} & 1 \\ 0 & 0 & 0 & \ldots & 0 & \lambda_{q+i} \end{Vmatrix}.$$

Therefore, the solution will have the form $\|n(t)\| = e^{Pt}\|n(0)\| = Me^{Jt}M^{-1}\|n(0)\|$, where $e^{Jt} = \text{diag}\{e^{J_0 t}, e^{J_1 t}, \ldots, e^{J_s t}\}$, $e^{J_0 t} = \text{diag}\{e^{\lambda_1 t}, e^{\lambda_2 t}, \ldots, e^{\lambda_q t}\}$, and

$$e^{J_i t} = e^{\lambda_{q+i} t} \begin{Vmatrix} 1 & t & \frac{t^2}{2!} & \cdots & \frac{t^{r_i-1}}{(r_i-1)!} \\ 0 & 1 & t & \cdots & \frac{t^{r_i-2}}{(r_i-2)!} \\ . & . & . & \cdots & . \\ 0 & 0 & 0 & \cdots & 1 \end{Vmatrix}.$$

This method of integration of kinetic equations (2) is general and allows one to obtain solutions for any probability matrix $P$. In specific calculations, two operations of matrix multiplication should be performed for each value of $t$, and the required calculation time, as can be shown, is proportional to $N^3$, which excludes the possibility of the real time modeling of the spectra of polyatomic molecules with $N > 10^3$. In addition, note that the computer calculation of the transformation matrix $M$ for such large values of $N$ is a nontrivial problem due to the errors in rounding off real numbers and the necessity of storage of large data bases.

The use of the approximate numerical methods for integrating kinetic equations (2), for example, based on the Eulerian piecewise linear functions method [29], seems the most promising. In this case, we obtain the point representation for the solution on the discrete grid $t_k$ in the form

$$\|n(t_{k+1})\| = \|n(t_k)\| + \Delta t P \|n(t_k)\|,$$

$$t_{k+1} = t_k + \Delta t.$$

Indeed, the rate of this iterative process is proportional to $N^2$ and the algorithm does not require the storage of additional matrices except the matrix $P$ itself. In addition, unlike the methods considered above, this method does not involve rather slow computer calculations of an exponential. Moreover, as the computer experiments showed, the numerical algorithm can be additionally optimized in rate by approximately two orders of magnitude due to the direct use of the matrix $P$ in the triangle form and also automatic exclusion of kinetic equations with zero solutions for the specified initial conditions. For this reason, the efficiency of this method proves to be satisfactory in practice. However, the most important advantage of this method is the fact that it yields the approximate solution of the system of kinetic equations with any matrix $P$, in particular, in the case of multiple eigenvalues, which, as we show below, is quite real.

Let us analyze the solutions of kinetic equations for some specific cases.

Consider first the simplest nontrivial example of the three-level system that has the ground state $S_0$ and two lowest excited electronic states $S_1$ and $S_2$ (without vibrational sublevels). In this case, the kinetic equations have the form

$$\frac{dn_2}{dt} = -w_2 n_2,$$

$$\frac{dn_1}{dt} = -w_1 n_1 + w_{21} n_2. \quad (7)$$

A simple analytic solution of the system (7) upon resonance excitation of the third level $\{n_2(0) = n_{20}, n_1(0) = 0\}$ can be easily obtained:

$$n_2(t) = n_{20} e^{-w_2 t}, \quad (8)$$

$$n_1(t) = n_{20}\frac{w_{21}}{w_2 - w_1}(e^{-w_1 t} - e^{-w_2 t}). \qquad (9)$$

If $w_1 = w_2$ (the multiple eigenvalue of the probability matrix), then $n_1(t) = n_{20} w_{21} e^{-w_1 t} t$.

Expression (8) describes a simple exponential law of the decrease in the population of the upper level $S_2$ (Fig. 1). Therefore, the decay rate of fluorescence $I_{20}(t)$ and $I_{21}(t)$ is characterized, according to (1) and (8), by a single quantity, namely, the total probability $w_2 = w_{20} + w_{21}$ of transitions from the $S_2$ state. The dependence $I_{10}(t) \sim n_1(t)$ is biexponential (9) and is already determined by two quantities, namely, the probabilities $w_2 = w_{20} + w_{21}$ and $w_1 = w_{10}$ of transitions from the $S_2$ and $S_1$ levels. Similar results were obtained in papers [23, 32].

Let us call attention to an important feature. The rate of the exponential increase in population $n_1(t)$ of the $S_1$ level and, hence, the intensity $I_{10}(t)$ of the $S_1 \longrightarrow S_0$ transition is $w_+ = \max(w_1, w_2)$, and the subsequent population decay occurs with the rate $w_- = \min(w_1, w_2)$. This follows from the fact that the sign of the difference $w_2 - w_1$ in expression (9) depends on the relation between the probabilities $w_1$ and $w_2$. One can see that already in this simplest three-level case the time dependence of the fluorescence intensity is determined not only by the probability of the given transition, but also by the probabilities of all the other transitions. For example, for $w_1 > w_2$, the rate of the exponential decay of fluorescence intensity $I_{10}(t)$ will be determined by the total probability $w_2 = w_{20} + w_{21}$ of transitions from the initially excited $S_2$ state rather than by the probability of the $S_1 \longrightarrow S_0$ transition. In the multi-level system containing vibrational levels as well, the situation is even more complicated (see below). Therefore, it is clear that a detailed and correct interpretation of the experiment, the determination of constants characterizing the lifetimes of energy levels, etc., can only be performed after the preliminary and detailed calculations of the spectra.

In the general case, when a system has many excited states, the time dependences of the populations are determined by expression (5) (or by a similar expression containing additional polynomial factors of $t$ in terms with identical total probabilities $w_i$). The first term is responsible for the transitions from the $i$th level to all the low-lying levels with the total probability $w_i = \sum_{j=1}^{i-1} w_{ij} + w_{i0}$, which characterizes the rate of depletion of this level. The rest of the terms contain in exponents the total probabilities of transitions from the levels lying above the $i$th level to this level. Therefore, the time dependence of the population of the $i$th level and, hence, of the intensity of transitions from this level is determined by the total probabilities $w_i$ of transitions both from this level and by the probabilities $w_k$ of transitions from all other levels lying above this level ($k > i$).

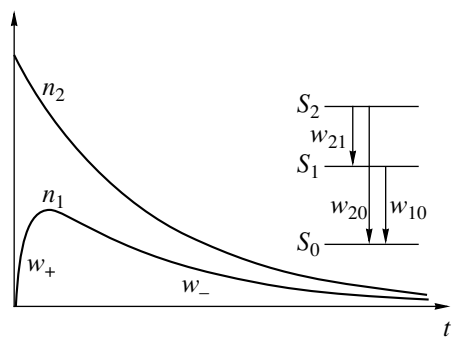

**Fig. 1.** Kinetic curves for a three-level system.

The total energy level diagram of a molecule can be separated into groups of vibrational levels belonging to different electronic states. Consider the ground 0 and the first excited 1 states of the molecule and their vibrational sublevels numerated by subscripts $f$ and $i$, respectively (these subscripts denote sets of vibrational quantum numbers). Let us calculate the total probabilities of vibronic transitions from the vibrational sublevels of the excited state to the ground state $w_{1,i} = \sum_f w_{1, i \to 0, f} =$

$$\sum_f \frac{64\pi^4 \nu_{1, i \to 0, f}^3}{3h\varepsilon_0 c^3} \mu_{1, i \to 0, f}^2, \quad \text{where} \quad \nu_{1, i \to 0, f} \quad \text{and}$$

$\mu_{1, i \to 0, f}$ are frequency and dipole moment of the vibronic transition, respectively.

In the Franck–Condon approximation, $\mu_{1, i \to 0, f} = \mu_{10} \int \Psi_i'(Q')\Psi_f(Q)dQ$ ($\mu_{10}$ is the dipole moment of the purely electronic transition), and the probabilities are determined by the overlap integrals for the wave functions $\Psi_i'$ and $\Psi_f$:

$$w_{1, i} = \frac{64\pi^4}{3h\varepsilon_0 c^3}\mu_{10}^2 \sum_f \nu_{1, i \to 0, f}^3 \langle i|f\rangle^2.$$

The terms containing the largest overlap integrals $\langle i|f\rangle$ will make the main contribution to the total probability. When the molecular geometry does not change significantly upon excitation, which is typical for polyatomic molecules, the overlap integrals containing functions with coinciding sets of quantum numbers $i$ and $f$ (or differing in the value of only one of them by unity) will be maximal. All the rest of the integrals will be substantially smaller. The frequencies of transitions with $i \approx f$ will differ from the purely electronic transition frequency $\nu_{10}$ by no more than the value of the order of a vibrational quantum (i.e., by several percent of the value of $\nu_{10}$). Therefore, we can assume on average with sufficient accuracy that $\nu_{1, i \to 0, f} \approx \nu_{10}$. Taking this into account and also that the vibrational wave func-

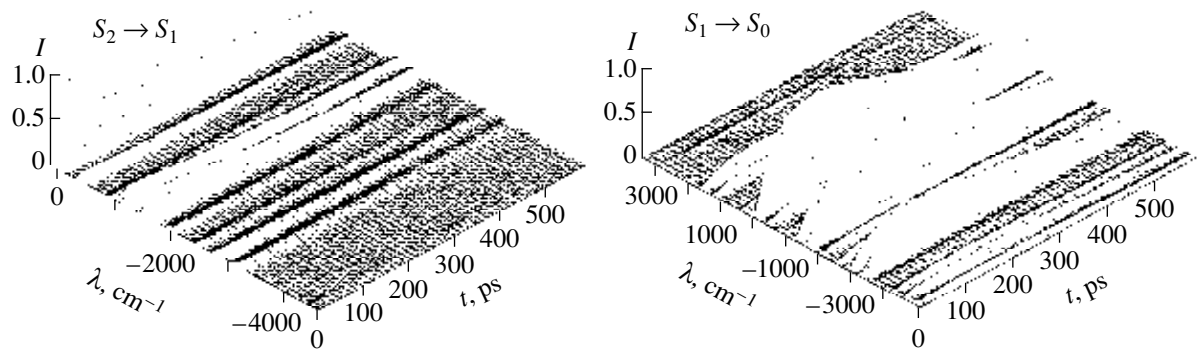

**Fig. 2.** Fluorescence spectra of the hexatriene model upon excitation of the purely electronic $S_2$ state.

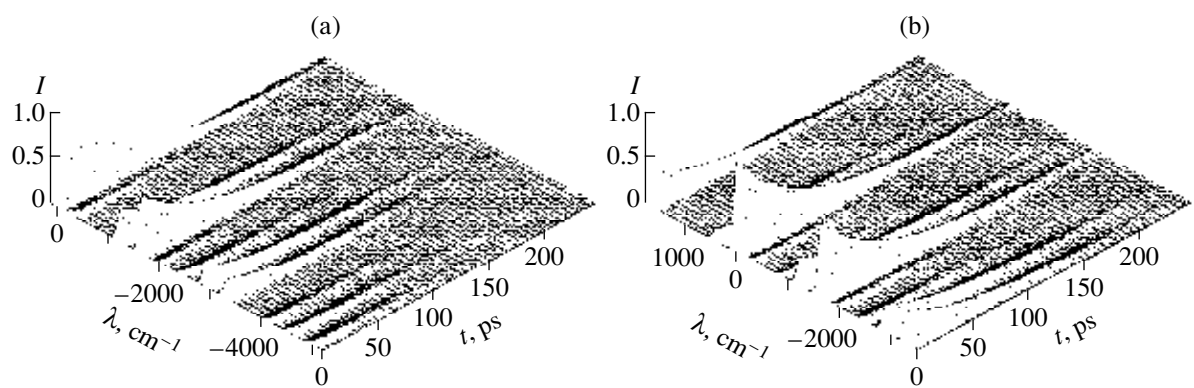

**Fig. 3.** The $S_1 \longrightarrow S_0$ fluorescence spectra of the octatetraene model upon excitation (a) of the purely electronic $S_1$ state and (b) the 1620-cm$^{-1}$ vibrational sublevel of the $S_1$ state.

tions form total orthonormal systems, we obtain for the total probability

$$w_{1,i} = \frac{64\pi^4}{3h\varepsilon_0 c^3}\nu_{10}^3 \mu_{10}^2 \sum_f \langle i|f\rangle^2 = \frac{64\pi^4}{3h\varepsilon_0 c^3}\nu_{10}^3 \mu_{10}^2. \quad (10)$$

One can see from this expression that the total probabilities (inverse lifetimes of the sublevels) for all vibrational sublevels of an electronic state are the same and are determined by the parameters $\mu_{10}$ and $\nu_{10}$ of the electronic transition. This conclusion is important and has been confirmed experimentally [23].

Consider now a system of three electronic states $S_0$, $S_1$, and $S_2$ with vibrational sublevels and vibronic transitions between them (the probabilities of vibrational transitions within the electronic states are assumed zero). Let the system be initially excited to the $S_2$. Then, the population of the sublevels of the intermediate state $S_1$ is described, according to (5), by the expression

$$n_{1,j} = \left(\sum_i b_{ij}\right) e^{-w_{1,j}t} - \sum_i b_{ij} e^{-w_{2,i}t},$$

because the population of the levels in the $S_1$ state increases only due to vibronic transitions from the $S_2$ state (the probabilities of vibrational transitions within the $S_1$ state are very small and neglected). Because the total probabilities for all the sublevels of the given electronic state are the same ($w_{1,j} = w_1$, $w_{2,i} = w_2$) (10), by neglecting vibrational transitions, the system of kinetic equations is separated into blocks, which correspond to different electronic states. The total probabilities are the same inside each block, and the equations are not connected with each other and are solved separately (but together with equations for other blocks). The solutions of these equations are proportional to the same sum of exponentials. Therefore, we obtain for the vibrational sublevels of the $S_1$ state:

$$n_{1,j} = \left(\sum_i b_{ij}\right)(e^{-w_1 t} - e^{-w_2 t}), \quad (11)$$

and the time dependences prove to be identical with an accuracy to the factor $\sum_i b_{ij}$ for all $j$, i.e., for all vibrational sublevels of the $S_1$ state.

From this, a result follows that directly affects the shape of the time-resolved spectrum: The time dynamics of all vibrational lines of the $S_1 \longrightarrow S_0$ electronic transition is the same, it is biexponential, and is determined by the total probabilities $w_1$ and $w_2$, while the intensities of the vibrational lines are determined by the coefficients $\sum_i b_{ij}$ and probabilities $w_{1, j \rightarrow 0, k}$ of the vibronic transitions.

When the probabilities of vibrational transitions (within the given electronic state) are not zero, expressions (10) and (11) are not valid in the general case and the dynamics of vibrational sublevels of the same electronic state becomes different. Therefore, the intensities of the vibrational lines of the $S_1 \longrightarrow S_0$ transition will change in time differently. The difference in the time dependences of the intensities of vibrational lines in the electronic spectrum is the criterion of a substantial role of vibrational transitions in the dynamics of the excited vibronic states of molecules and shows that the probabilities of vibrational transitions are comparable with those of vibronic transitions.

Upon population of high vibrational levels (overtones and combination frequencies), a situation can arise in which the relaxation will occur via the vibrational states, with the probabilities of transitions between them being close or identical (for example, for the most intense vibrational transitions accompanied by a change in one of the quantum numbers by unity). In this case, the total probabilities (including also the electronic component) will be the same for the nonzero probabilities of vibrational transitions within electronic states as well. In this situation, the system of kinetic equations cannot be separated into blocks, as in the case of negligibly small probabilities of vibrational transitions, and its matrix will have multiple eigenvalues, which requires the use of general methods for solving kinetic equations in calculations of the spectra.

One can expect that this will be most typical in the analysis of the time-resolved fluorescence spectra of polyatomic molecules upon selective (resonance) excitation of vibronic states. Therefore, the iterative method of numerical integration of kinetic equations (2) described above appears the most adequate. We developed a convenient calculation algorithm and a software for the construction of three-dimensional spectra and performing computer experiments with real molecules. Figures 2 and 3 illustrate the spectra calculated upon selective excitation of different vibronic states of models of hexatriene and octatetraene molecules. The results of a series of computer experiments and their detailed discussion will be reported elsewhere.

## CONCLUSIONS

Thus, the problem of modeling time-resolved vibronic spectra of polyatomic molecules can be solved using the parametric method and molecular models, which were developed earlier for the calculation and prediction of conventional absorption and fluorescence spectra of polyatomic molecules. All the required matrix elements can be calculated by the methods developed. The method of numerical integration of a system of many ($N > 10^3$) kinetic equations is the most efficient for determining the time dependences of the populations of vibronic levels.

We realized the parametric method of modeling the time-resolved vibronic spectra of polyatomic molecules in the form of special software for a PC. The three-dimensional spectra were calculated for molecules of polyenes and diphenylpolyenes under different excitation conditions and for different parameters of molecular models. The calculations showed the possibility of performing real time computer experiments with polyatomic molecules containing several tens of atoms.

The analysis of kinetic equations showed, in particular, that the time dependence of the intensity of vibrational lines (which is many-exponential in the general case) can be used for estimating the role of vibrational transitions in the dynamics of excited vibronic states. The direct calculation and modeling of these spectra allow one to separate radiative and nonradiative contributions to the probabilities of the corresponding transitions, which is important in the development of the models of intermolecular interactions and the theory of nonradiative transitions [21].

We restricted ourselves to the processes caused by the vibronic transitions occurring within the same molecular isomer. In the general case, processes of optical excitation and relaxation can result in the transitions to other molecular isomers, which is of great interest, in particular, for photochemistry. We intend to study these processes in the future. After the development of the methods for calculating probabilities of vibronic transitions between isomers [14], this method can be easily extended to the general case. Note that the simulation of three-dimensional spectra neglecting the transitions between isomers and their comparison with experiments can already give the information that will be useful for the development of the theory of optically induced transitions between isomers of polyatomic molecules.

## REFERENCES

1. J. C. Brown, J. M. Hayes, J. A. Warren, and G. J. Small, *Laser in Chemical Analysis*, Ed. by G. M. Hieftje, J. M. Travis, and F. E. Lytie (The Humana Press, New York, 1981).
2. J. A. Warren, J. M. Hayes, and G. J. Small, Anal. Chem. **54**, 138 (1982).
3. D. H. Levy, Annu. Rev. Phys. Chem. **31**, 197 (1980).
4. A. Amirav, U. Even, and J. Jortner, Chem. Phys. **51** (1–2), 31 (1980).
5. S. M. Beck, D. E. Powers, J. B. Hopkins, and R. E. Smally, Chem. Phys. **74**, 43 (1981).

6. U. Even, J. Magen, J. Jortner, and C. Levanon, J. Am. Chem. Soc. **103**, 4583 (1981).
7. R. D. Cordon and J. M. Hollas, J. Mol. Struct. **293**, 193 (1993); J. Chem. Phys. **99**, 3380 (1993).
8. K. B. Eisenthal, Annu. Rev. Phys. Chem. **28**, 207 (1977).
9. J. L. Kinsey, Annu. Rev. Phys. Chem. **28**, 349 (1977).
10. A. H. Zewail, Adv. Chem. Phys. **101**, 892 (1997).
11. *Femtochemistry: Ultrafast Chemical and Physical Processes in Molecular Systems*, Ed. by M. Chergui (World Scientific, Singapore, 1996).
12. D. N. Shigorin, G. A. Val'kova, E. A. Gastilovich, *et al.*, *Excited Electronic States of Polyatomic Molecules and Processes of Their Decontamination* (Nauka, Moscow, 1993).
13. L. A. Gribov, V. I. Baranov, and D. Yu. Zelentsov, *Electronic-Vibrational Spectra of Polyatomic Molecules. Theory and Computing Methods* (Nauka, Moscow, 1997).
14. L. A. Gribov and I. V. Maslov, J. Mol. Struct. **521** (1–3), 107 (2000).
15. L. A. Gribov and I. V. Maslov, Zh. Fiz. Khim. **74**, 441 (2000).
16. V. I. Baranov and D. Yu. Zelent'sov, J. Mol. Struct. **328** (1), 179 (1994).
17. V. I. Baranov, L. A. Gribov, and D. Yu. Zelent'sov, J. Mol. Struct. **328** (1), 189 (1994).
18. V. I. Baranov and D. Yu. Zelent'sov, J. Mol. Struct. **328** (1), 199 (1994).
19. V. I. Baranov, L. A. Gribov, and D. Yu. Zelent'sov, J. Mol. Struct. **376**, 475 (1996).
20. V. I. Baranov, L. A. Gribov, V. O. Djenjer, and D. Yu. Zelent'sov, J. Mol. Struct. **407** (2–3), 177 (1997).
21. V. I. Baranov and L. A. Gribov, Zh. Prikl. Spektrosk. **67** (3) (2000).
22. L. A. Gribov and V. I. Baranov, Opt. Spektrosk. **85**, 46 (1998) [Opt. Spectrosc. **85**, 41 (1998)].
23. J. A. Syage, P. M. Felker, and A. H. Zewail, J. Chem. Phys. **81**, 4706 (1984).
24. G. A. Zalesskaya, D. L. Yakovlev, E. G. Sambor, and D. I. Baranovskiĭ, Opt. Spektrosk. **84**, 733 (1998) [Opt. Spectrosc. **84**, 658 (1998)].
25. V. I. Baranov and A. N. Solov'ev, Zh. Strukt. Khim. **41**, 368 (2000).
26. L. A. Gribov and V. A. Dement'ev, *Computational Methods and Algorithms in Theory of Molecular Vibrational Spectrum* (Nauka, Moscow, 1981).
27. L. A. Gribov, V. A. Dement'ev, and O. V. Novoselova, *Interpreted Vibrational Spectra of Hydrocarbons with Isolated and Conjugated Multiple Bonds* (Nauka, Moscow, 1987).
28. V. I. Baranov, F. A. Savin, and L. A. Gribov, *Calculating Programs of Electronic-Vibrational Spectra of Polyatomic Molecules* (Nauka, Moscow, 1983).
29. V. I. Arnol'd, *Ordinary Differential Equations* (Nauka, Moscow, 1975; MIT Press, Cambridge, 1973).
30. F. R. Gantmacher, *The Theory of Matrices* (Fizmatgiz, Moscow, 1967, 3rd ed.; Chelsea, New York, 1959).
31. E. A. Coddington and N. Levinson, *Theory of Ordinary Differential Equations* (McGraw-Hill, New York, 1955; Inostrannaya Literatura, Moscow, 1958).
32. J. S. Baskin, L. Banares, S. Pedersen, and A. H. Zewail, J. Phys. Chem. **100**, 11920 (1996).

*Translated by M. Sapozhnikov*